\documentclass[pdflatex,sn-nature,iicol]{sn-jnl}
\unnumbered
\pdfoutput=1
\usepackage{graphicx}
\usepackage{multirow}
\usepackage{amsmath,amssymb,amsfonts}
\usepackage{amsthm}
\usepackage{mathrsfs}
\usepackage[title]{appendix}
\usepackage{xcolor}
\usepackage{textcomp}
\usepackage{manyfoot}
\usepackage{booktabs}
\usepackage{algorithm}
\usepackage{algorithmicx}
\usepackage{algpseudocode}
\usepackage{listings}
\usepackage{anyfontsize}
\begin{document}

\title{Single electron interference and capacitive edge mode coupling generates $\phi_0/2$ flux periodicity in Fabry-P\'erot interferometers}

\author[1,2]{\fnm{S.} \sur{Liang}}

\author[1,2]{\fnm{J.} \sur{Nakamura}}

\author[2,3]{\fnm{G. C.} \sur{Gardner}}

\author*[1,2,3,4,5]{\fnm{M. J.} \sur{Manfra}}\email{mmanfra@purdue.edu}

\affil[1]{Department of Physics and Astronomy, Purdue University, West Lafayette, IN, USA}

\affil[2]{Birck Nanotechnology Center, Purdue University, West Lafayette, IN, USA}

\affil[3]{Microsoft Quantum Lab West Lafayette, West Lafayette, IN, USA}

\affil[4]{Elmore Family School of Electrical and Computer Engineering, Purdue University, West Lafayette, IN, USA}

\affil[5]{School of Materials Engineering, Purdue University, West Lafayette, IN, USA}

\abstract{Experimental observations of flux periodicity $\phi_{0}/2$, where $\phi_0=h/e$, for interference of the outermost edge mode in the integer quantum Hall regime have been attributed to an exotic electron pairing mechanism. We present measurements of an AlGaAs/GaAs Fabry-P\'erot interferometer operated in the integer quantum Hall regime for filling factors $1\leq \nu \leq 3$ that has been designed to simultaneously express measurable bulk-edge and edge-edge couplings. At integer fillings $\nu=2$ and $\nu=3$, we observe interference with flux periodicity $\phi_{0}/2$ for the outermost edge mode. Our analysis indicates that the periodicity $\phi_0/2$ is not driven by electron pairing but is the result of capacitive coupling between isolated edge modes and the interfering outer edge. The interfering unit of charge for the outermost edge mode at $\nu=2$ and $\nu=3$ was determined to be $e^*=1$, where the effective charge $e^*$ is normalized to the charge of an electron. Our measurements demonstrate that the magnitude of the interfering charge can be determined {\it in operando} in a Fabry-P{\'e}rot interferometer.}

\maketitle

\section{Introduction}\label{sec:level1}
Electrons confined to two spatial dimensions and subjected to a strong perpendicular magnetic field exhibit quantum Hall effects~\cite{Tsui82,Stormer99}. Charge transport is governed by one-dimensional chiral edge modes~\cite{Halperin82,wen90,Chklovskii92,Chklovskii1993,Von11}, enabling coherent transport and interferometry. Fabry-P\'erot interferometers serve as powerful probes for studying quasiparticle charge, anyonic braiding statistics, and Coulomb interactions between localized charged excitations and gapless edge modes. Fabry-P\'erot interferometers have been extensively explored in both theoretical~\cite{Kivelson89,Kivelson90,Gefen93,chamon97,Kim06,Bonderson06,Stern06,Halperin07,Stern08,Bishara2008,Bishara2009,Halperin2011,Rosenow12,Levkivskyi12,Feldman14,Keyserlingk15,Sahasrabudhe2018,Frigeri2020,Rosenow2020,Carrega21,Feldman21,Feldman22} and experimental works~\cite{Deviatov08,Roulleau08,Zhang09,Mcclure09,Ofek10,Mcclure12,Kou12,Willett09,Willett13,Willett19,Jimmy19,Jimmy20,Jimmy22,Jimmy2023,Ronen21,Deprez21,Zhao22}. While much recent effort has focused on interrogation of braiding statistics in the fractional quantum Hall regime, Fabry-P\'erot interferometers also expose interesting phenomena in the integer quantum Hall regime. Unexpectedly, experiments~\cite{Choi15,sivan16,Sivan18} have revealed a halved magnetic flux period of $\phi_0/{2}$ and a halved voltage period while interfering the outermost edge mode for filling factors $2.5<\nu<5$. Here $\phi_0\equiv\frac{h}{e}$ is the quantum of magnetic flux. The authors of~\cite{Choi15,sivan16,Sivan18} proposed that the periodicity of $\phi_0/2$ was due to the pairing of electrons, resulting in an interference charge of $e^*=2$, where $e^*$ is normalized to the magnitude of the charge of a single electron $|e|$. However, the mechanism that generates the putative pairing was not established. Subsequent theoretical work~\cite{Frigeri2019} argued that the period halving may be understood as an effect not of electron pairing but of electrostatic interaction between multiple independent edge channels. A recent experiment in a graphene-based Fabry-P{\'e}rot interferometer concluded that edge-edge coupling and charge quantization were enough to generate the observed period halving without invoking electron pairing~\cite{werkmeister2024}.

In a quantum Hall Fabry-P\'erot interferometer, two quantum point contacts (QPCs) are used to partially reflect incident edge modes. Interference is generated by coherent transport within the interferometer chamber. In the limit of weak backscattering and negligible Coulomb coupling, the interference phase evolves as~\cite{chamon97,Halperin2011,Feldman21,Feldman22},

\begin{equation} \label{phase of pure AB interference}
\frac{\theta}{2\pi} = e^*\frac{AB}{\phi_0}+N_{L}\frac{\theta_a}{2\pi}
\end{equation}
Here $B$ is the magnetic field, $A$ is the area enclosed by the interference path, $e^*$ denotes the quasiparticle charge normalized to the elementary electron charge
$|e|$, $N_L$ is the number of quasiparticles localized within the device, and $\theta_a$ is the statistical phase of the quasiparticles. Eqn.~\ref{phase of pure AB interference} incorporates both the Aharonov-Bohm phase and the statistical (anyonic) phase contributions to the interference phase. For integer quantum Hall states, the discrete change in phase associated with addition or subtraction of a localized electron is an unobservable factor of $2\pi$. Hence, the variation of the interference phase in integer states in the absence of Coulombic couplings can be expressed as ${\delta \theta}/{2\pi} = {(A\delta B+B\delta A)}/{\phi_0}$, where $\delta A=\alpha \delta V_{PG}$ is the variation in area due to the modulation of the plunger gate and $\alpha$ is the constant of proportionality relating the change in the interferometer area to the variation of the plunger gate voltage. The expected magnetic field and plunger gate voltage periods are:

\begin{equation} \label{AB Bperiod}
\Delta B=\frac{\phi_0}{A}
\end{equation}
\begin{equation} \label{AB Vperiod}
\Delta V_{PG}=\frac{\phi_0}{\alpha B}
\end{equation}
Eqn.~\ref{phase of pure AB interference} neglects two important couplings active in real interferometers: bulk-edge coupling, which may alter the area of the interference path when the charge configuration in the interior of the interferometer changes, and capacitive edge-edge coupling that may contribute when isolated edge modes are present at the boundary of the quantum Hall liquid. At integer filling factors greater than $\nu=1$ and in certain fractional states, multiple edge modes are typically present. In these cases, we must consider the capacitive interaction between multiple edge modes. 

Theoretical work has extensively examined a minimal capacitive model to describe the energetics of a Fabry-P\'erot interferometer~\cite{Halperin07,Keyserlingk15,Halperin2011,Frigeri2019,Feldman21,Feldman22}. As a specific example, we consider an interferometer in which the bulk filling factor is $\nu=2$ bounded by two circulating edge modes. For small variations in the charge in the interior of the interferometer and the interferometer area, the total electrostatic energy of the system may be expressed to quadratic order as:
\begin{eqnarray}
E_{tot}=&&\frac{K_1}{2} \delta q_1^2+\frac{K_2}{2} \delta q_2^2+\frac{K_L}{2} \delta q_b^2+K_{1L}\delta q_b \delta q_1\nonumber\\
&&+K_{2L}\delta q_b \delta q_2+K_{12}\delta q_1 \delta q_2 \label{total energy}
\end{eqnarray}
The variation in charge in the interior of the interferometer is represented by $\delta q_b$, the variation of charge on the outer edge is $\delta q_1$, and the variation of charge on the inner edge mode is $\delta q_2$. A set of effective interactions parameterize the total electrostatic energy. The stiffness of the edge modes, $K_1$ and $K_2$, represents the energy cost associated with the variation of the area of the edge channels for the outer edge and the inner edge, respectively; bulk-edge couplings, $K_{1L}$ and $K_{2L}$, quantify the electrostatic coupling between charge in the interior of the interferometer and each of the two edge modes, and the edge-edge coupling, $K_{12}$, captures the electrostatic (capacitive) interaction between the two edge modes. 

The variations of charge in Eqn.~\ref{total energy} may be further specified as:
\begin{equation}
    \delta q_b=e^*N_L+2\frac{\bar AB}{\phi_0}-\gamma \Delta V_{PG}
\end{equation}
\begin{equation}
    \delta q_1=(N_1-\frac{\bar AB}{\phi_0})
    \end{equation}
 \begin{equation}
    \delta q_2=(N_2-\frac{\bar AB}{\phi_0})  
 \end{equation}
Here $\bar{A}$ is the average area of the interferometer that does not include variations $\delta A$ due to bulk-edge and/or edge-edge coupling. $N_1$ is the charge in N=0 spin-down Landau level and $N_2$ is the charge in the N=0 spin-up Landau level that forms the quantum Hall condensate in the interior of the interferometer, not including localized excitations. $\gamma$ is the proportionality constant that relates the change in the voltage of the plunger, $\Delta V_{PG}$, needed to induce the charge $|e|$ on the interferometer. In a configuration where the QPCs are set to partially transmit the outer mode while fully reflecting the inner mode, the charge dynamics is constrained. Charge variations in the interior of the interferometer and on the inner isolated edge are quantized to integer values because the inner edge is effectively disconnected from the reservoir of charge in the ohmic leads.

Minimization of the total energy with respect to the variations of the charge on the interfering outer edge is achieved by setting $\frac{\partial E_{tot}}{\partial{(\delta q_1)}}=0$. This condition leads to the following relation for modifications of charge in the interior of the interferometer (bulk and isolated inner edge mode) and on the outer interfering edge mode: 
\begin{equation}
\delta q_1=-(K_{1L}/K_1) \delta q_b-(K_{12}/K_1) \delta q_2.   
\end{equation}
Using the established relationships between $\delta q_i$'s, the expression for the total interference phase (Eqn.~\ref{phase of pure AB interference}) may be rewritten as:
\begin{eqnarray}
\frac{\theta}{2\pi}=&&e^*\frac{\bar{A}B}{\phi_0}-e^*\frac{K_{1L}}{K_1}(e^*N_L+2\frac{\bar{A}B}{\phi_0}-\gamma \Delta V_{PG})\nonumber\\
&&{} - e^*\frac{K_{12}}{K_1}(N_2-\frac{\bar{A}B}{\phi_0})
\label{equation total phase variation for edge-edge coupling}
\end{eqnarray}
The normalized charge of the interfering particles is retained as $e^*$ to maintain generality even though we may expect $e^*=1$ for the integer states. For simplicity, we neglect the anyonic phase term $N_{L}\frac{\theta_a}{2\pi}$ in Eqn.~\ref{phase of pure AB interference} as we do not expect anyonic statistics at the integer states $\nu=2$ and $\nu=3$.
In Eqn.~\ref{equation total phase variation for edge-edge coupling}, the second term on the right-hand side reflects the contribution from the bulk-edge coupling, while the third term is the phase variation attributable to edge-edge coupling. 

The interplay between bulk-edge and edge-edge couplings significantly modifies the interference. Eqn.~\ref{equation total phase variation for edge-edge coupling} is generally applicable and does not rely on a particular regime of interferometer operation. For interferometer operation in incompressible regimes, where the number of localized quasiparticles is fixed such that $\delta N_L=0$, modifications to the interference periods induced by these couplings can be expressed in a simple form. For an incompressible regime at $\nu=2$, Eqn.~\ref{equation total phase variation for edge-edge coupling} implies magnetic field and gate voltage periods given by the following expressions:

\begin{equation} \label{Bperiod in the incompressible region for edge-edge coupling}
\Delta B=\frac{\phi_0}{e^*(1-2\frac{K_{1L}}{K_1}+\frac{K_{12}}{K_1})\bar{A}}
\end{equation}
\begin{equation} \label{Vperiod in the incompressible region for edge-edge coupling}
\Delta V_{PG}=\frac{1}{e^*}((1-2\frac{K_{1L}}{K_1}+\frac{K_{12}}{K_1}) \frac{\alpha B}{\phi_0}+\gamma\frac{K_{1L}}{K_1})^{-1}
\end{equation}

These particular expressions for the magnetic field and plunger gate voltage periods are valid when $N_2$ is constant and $\delta N_L=0$ such that the interference phase evolves continuously. Even within an incompressible region, if the number of localized quasiparticles varies, discrete phase slips can also be observed, as predicted in theoretical studies~\cite{Halperin2011,Keyserlingk15,Frigeri2019,Feldman21}. Between these discrete phase slips, $N_L$ remains fixed and the interference phase evolves continuously. The periodicity of phase evolution between phase slips is described by Eqn.~\ref{Bperiod in the incompressible region for edge-edge coupling} and Eqn.~\ref{Vperiod in the incompressible region for edge-edge coupling} and may be determined experimentally through Fourier decomposition of the interference data. The interference charge in the integer quantum Hall regime can be determined by analysis of the discrete phase slips, while the ratios $K_{1L}/K_1$ and $K_{12}/K_1$ can be accurately determined by analyzing the interference periods in compressible and incompressible operating regimes, and through analysis of differential conductance measurements~\cite{Jimmy22,Jimmy2023}. The effective area of the interferometer and the proportionality constants can be extracted by measuring the differential conductance in the tunneling regime at $B=0$~T and the interference at $\nu=1$~\cite{Jimmy19,Jimmy20,Jimmy22,Jimmy2023}.

If interference occurs in the compressible regime, the analysis of the expected magnetic field period which we used to evaluate $K_{12}/K_1$ is modified slightly. In the compressible regime, the total charge density in the interior of the interferometer remains fixed ($\partial (\delta q_b)/\partial B=0$) when the magnetic field is varied so that the flux changes by $\phi_0$. Examination of Eqn.~\ref{equation total phase variation for edge-edge coupling} indicates that the magnetic field period is modified in the compressible regime to be the following:
\begin{equation}
    \Delta B = \frac{\phi_0}{\bar{A}(1+\frac{K_{12}}{K_1})}
\end{equation}

Here we report on an experiment designed to determine the magnitude of the interfering charge and understand the origin of the unusual $\phi_0/2$ periodicity for the interference of the outermost edge mode at $\nu=2$ and $\nu=3$ in an AlGaAs/GaAs Fabry-P{\'e}rot interferometer. Toward this end, we designed and examined an interferometer that expresses measurable bulk-edge coupling and edge-edge coupling to highlight the important role played by these Coulomb coupling mechanisms in determining interferometer periodicity and phase evolution in the integer quantum Hall regime. We describe and apply a method to extract {\it in operando} the effective interference charge, the edge-edge coupling constant, and the bulk-edge coupling constant~\cite{Jimmy22,Jimmy2023}. The magnitude of the couplings extracted by this analysis quantitatively explain the observed $\phi_0/2$ periodicity and constrain the possible magnitude of the interfering charge. The central findings of our study can be summarized as: 1) the observed periodicity $\phi_0/2$ is quantitatively explained by capacitive coupling to isolated inner edge modes, and 2) the magnitude of discrete phase slips at $\nu=2$ and $\nu=3$ for interference of the outer edge mode indicate that the interfering charges are single electrons, ruling out for our experiments an effective charge $e^*=2$ postulated previous experimental works~\cite{Choi15,sivan16,Sivan18}.

\section{Results}\label{sec:level2}
\subsection{Device Design}\label{subsec:level1}
\begin{figure}
\centering
\includegraphics[width=0.5\textwidth]{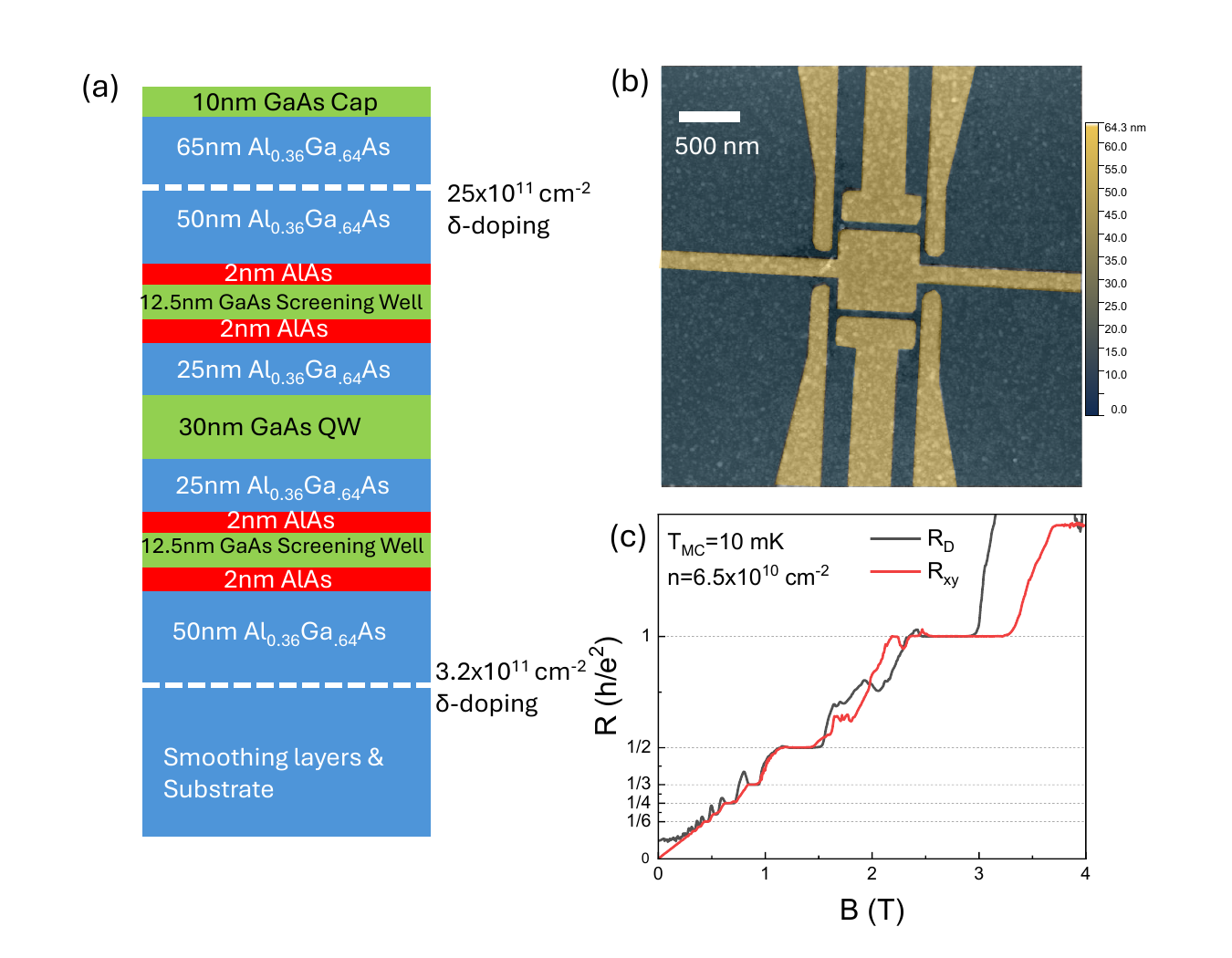}
\caption{\label{Fig:1} \textbf{Device Design} (a) Schematic of our GaAs/AlGaAs heterostructure consisting of a primary GaAs quantum well flanked by ancillary GaAs screening wells. (b) False color Atomic force microscopy image of the interferometer studied in this work. Yellow regions are the metallic gates that define the interference path with a lithographic area 0.58~$\mu$m$^2$. (c) Bulk $R_{xy}$ and diagonal resistance $R_D$ across the interferometer demonstrating overlapping plateaux at integer quantum Hall states. The QPCs are biased at $V_{QPC}=-0.9$~V and the plunger gate is biased at $V_{PG}=-0.7$~V, just past depletion. The top gate in the center of the interferometer is grounded during all measurements.
}
\end{figure}

Our device was fabricated on the AlGaAs/GaAs heterostructure~\cite{Manfra14,Gardner16,Jimmy19} shown in Fig.~\ref{Fig:1}\textcolor{blue}{a}. Our heterostructure design incorporates three GaAs wells populated with electrons: a 30~nm wide primary quantum well flanked by two additional 12.5~nm wells situated on either side of the primary well separated by 25~nm AlGaAs spacer layers. The 2DEG located in the primary GaAs quantum well has an electron density $n_s=0.65\times10^{11}$~cm$^{-2}$. The additional populated quantum wells serve to partially screen long-range Coulomb interactions, allowing the interferometer to operate in an intermediate regime that exhibits features associated with both Aharonov-Bohm interference and Coulomb-dominated interference. Electron beam and optical lithography techniques were employed in the fabrication process. In Fig.~\ref{Fig:1}\textcolor{blue}{b}, an atomic force microscopy image of the interferometer is presented. QPCs form narrow constrictions and a pair of plunger gates define the interference path. The QPCs are negatively biased to adjust the transmission of the individual edge modes, while the plunger gates are biased just past the depletion point to delineate the interference path. The central top gate was grounded during the course of this experiment. In Fig.~\ref{Fig:1}\textcolor{blue}{c}, the Hall resistance, $R_{xy}$, and the diagonal resistance across the interferometer, $R_D$, as a function of magnetic field are plotted with the QPCs and plunger gates biased just past depletion. Parallel conduction through the ancillary wells is suppressed by negatively biasing gates near the ohmic contacts~\cite{Jimmy19,Jimmy20,Jimmy22,Jimmy2023}.

\subsection{Measurements at \texorpdfstring{$B=0$ and $\nu=1$}{Lg}}\label{subsec:level2}

\begin{figure}
\centering
\includegraphics[width=0.5\textwidth]{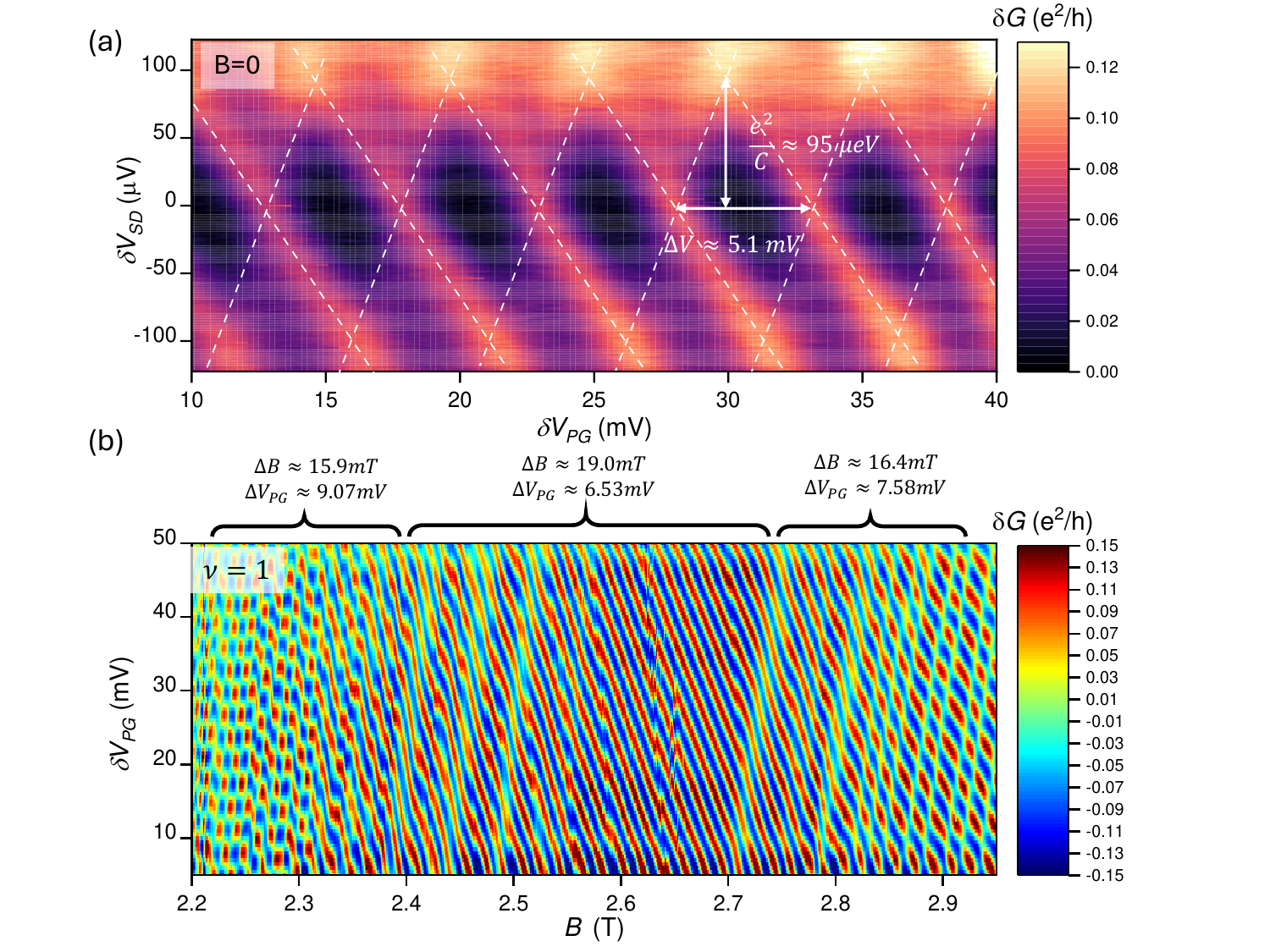}
\caption{\label{Fig:2} \textbf{Measurements at $B=0$ and $\nu=1$} (a) Differential conductance measurements at zero magnetic field in the Coulomb blockade regime. The height of the diamonds gives the charging energy of $\frac{e^2}{C}\approx95$~$\mu$eV. Additionally, the spacing between crossings at zero bias yields $\frac{1}{\gamma}\approx5.1$~mV. (b) Interferometer conductance oscillations versus magnetic field and gate voltage at $\nu=1$. The magnetic field oscillation period is larger near the plateau center, indicating an incompressible bulk and moderate bulk-edge coupling. In the compressible regions at high and low fields, the interference oscillation period yields an effective area of $A\approx0.26$~$\mu$m$^2$ and $\alpha=\frac{\partial A}{\partial V_{PG}}\approx0.198$~$\mu$m$^2$V$^{-1}$.
}
\end{figure}

Fig.~\ref{Fig:2}\textcolor{blue}{a} illustrates the variation in conductance, $\delta G$, as a function of the voltage variation of the plunger gate, $\delta V_{PG}$, and the variation of the DC source-drain bias, $\delta V_{SD}$, in the tunneling regime at $B=0$~T. The height of the Coulomb diamonds indicates that the charging energy is $E_C\approx 95$~$\mu$eV. The spacing between crossings at zero DC bias yields $1/\gamma \approx 5.1$~mV. In Fig.~\ref{Fig:2}\textcolor{blue}{b}, the conductance variation, $\delta G$, (with a smooth background subtracted) is plotted against magnetic field $B$ and plunger gate voltage variation, $\delta V_{PG}$, at the integer quantum Hall state $\nu=1$. The QPCs have been set to approximately 90~$\%$ transmission. Note that $\delta V_{PG}$ is relative to -0.7~V in all measurements. At the center of the $\nu=1$ plateau ($B\approx 2.6$~T), the magnetic field period is $\Delta B \approx 19.0$~mT. The period in the center of the plateau is longer than that observed in the low and high field flanks, indicating that the chemical potential lies in the bulk gap and the state is incompressible~\cite{Halperin2011,Rosenow2020,Feldman21}. At higher and lower fields, the magnetic field period is shorter, indicating that the localized electron number varies by 1 when the flux changes by $\phi_0$, returning the magnetic field period to the base value of $\phi_0/A$. The change in periodicity in the higher and lower fields is consistent with a transition from an incompressible regime in the center of the plateau to compressible regimes on the flanks~\cite{Halperin2011,Rosenow2020,Feldman21,Jimmy20,Jimmy22}. The average magnetic field period at the higher and lower fields, $\Delta B=16.2$~mT, yields an effective area of the interferometer $A=\phi_0/\Delta B\approx 0.26$~$\mu$m$^2$. The gate voltage oscillation period at $B=2.3$~T is $\Delta V_{PG}=9.07$~mV, which yields $\partial A /\partial V_{PG}=\phi_0/(B \Delta V_{PG})\approx 0.198$~$\mu$m$^2$V$^{-1}$. The ratio of periods in the incompressible regime and in the compressible regime can be used to estimate $K_{IL}/K_I \approx 0.16$, indicating a relatively weak bulk-edge coupling at $\nu=1$ for this device consistent with the observed Aharonov-Bohm interference pattern~\cite{Jimmy22,Jimmy2023, Halperin2011,Keyserlingk15,Frigeri2019,Rosenow2020,Feldman21}.
\subsection{Interference at \texorpdfstring{$\nu=2$}{Lg}}\label{subsec:level3}

\begin{figure*}
\centering
\includegraphics[width=0.8\textwidth]{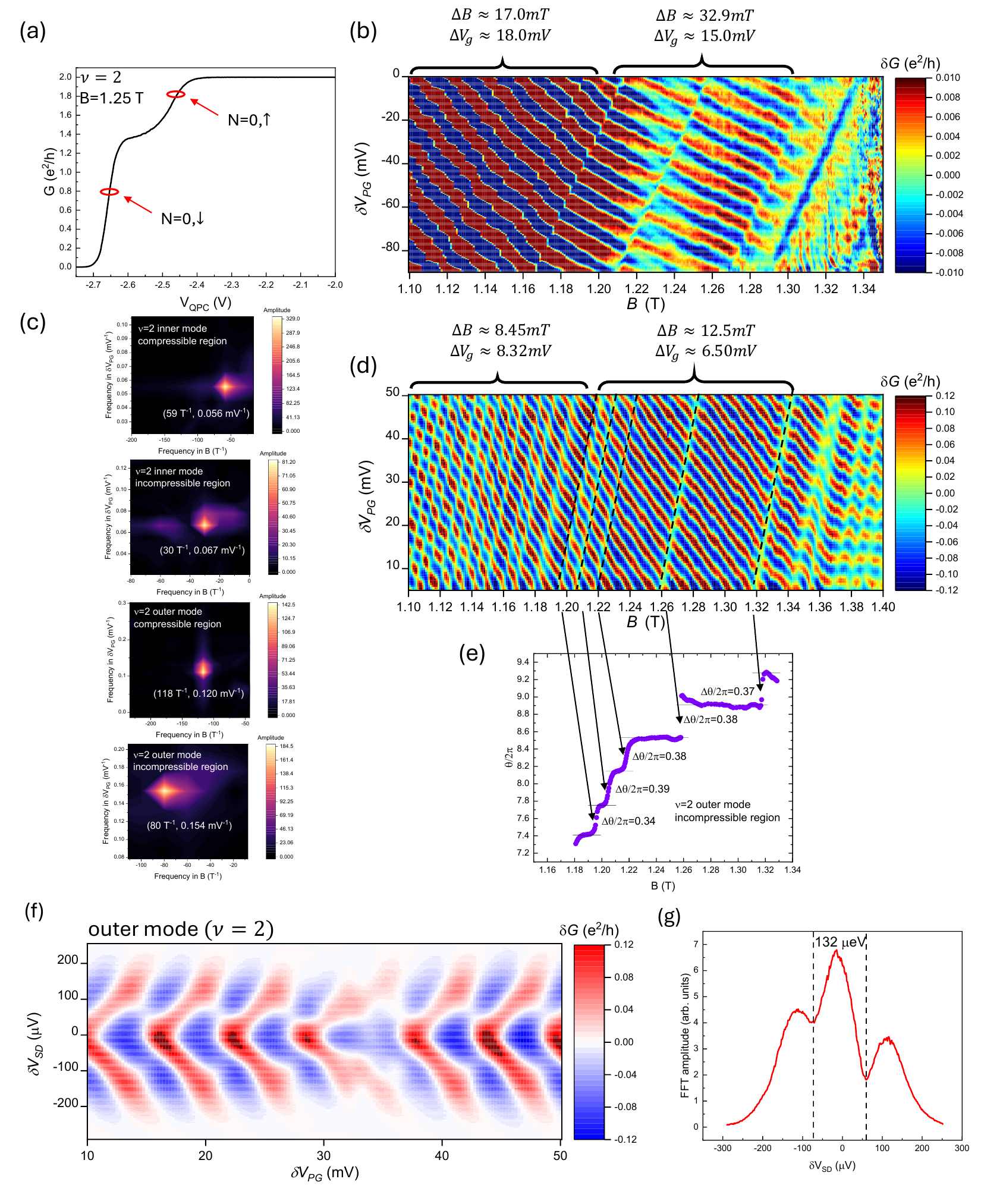}
\caption{\label{Fig:3} \textbf{Interference at $\nu=2$} (a) QPC conductance versus gate voltage at $\nu=2$. A robust conductance plateau at $G=2\frac{e^2}{h}$ is observed, indicating full transmission of two edge modes of the N=0 Landau level. However, a sharp plateau at $G=\frac{e^2}{h}$ is not evident, likely due to the smallness of the spin gap at $B=1.25$~T. Red circles indicate the QPC settings for partial transmission of the inner mode and outer mode of the N=0 Landau level. (b) Interference while partially transmitting the inner edge mode (spin-up N=0 Landau level). (c) 2D FFTs of interference for the inner mode and the outer mode in the compressible and incompressible regimes at $\nu=2$. (d) Interference while partially transmitting the outer edge (spin-down N=0 Landau level). For interference of both the inner and outer modes, the data display a larger magnetic field period near the plateau center, indicating an incompressible bulk. At lower fields, a smaller period accompanied by numerous phase modulations is observed, suggesting a compressible bulk. Black dashed lines in panel (d) highlight the discrete phase slips within the incompressible region when interfering the outer edge mode at $\nu=2$. (f) Differential conductance versus DC source-drain bias and plunger gate voltage. (g) FFT amplitude of the differential conductance vs. DC source-drain bias. The FFT amplitude minimum corresponds to the nodes in the interference, allowing extraction of $\Delta V_{SD1}\approx132$~$\mu$V.  
}
\end{figure*}

Our primary objective is to determine the interfering charge and to understand the origin of the unusual flux periodicity $\phi_0/2$ for interference of the outermost edge mode at $\nu=2$ and $\nu=3$ that has been attributed to electron pairing in previous experiments. Setting the magnetic field to $B=1.25$~T places the interferometer at $\nu=2$. The conductance of an individual QPC at $\nu=2$ is shown in Fig.~\ref{Fig:3}\textcolor{blue}{a}. The complete transmission of two edge modes is revealed by the quantized plateau at $G=2e^2/h$. The absence of a clear plateau $G=e^2/h$ suggests coupling between the inner and outer edges, likely associated with a small spin gap at $B=1.25$~T. We first set the QPCs to partially reflect the inner edge mode. We eventually interfere the outer edge mode by applying more negative bias. The operating points are indicated by red circles in Fig.~\ref{Fig:3}\textcolor{blue}{a}. 

Fig.~\ref{Fig:3}\textcolor{blue}{b} displays interference when weakly backscattering the {\it inner} mode and fully transmitting the outer mode. On the low-field flank of $\nu=2$ ($B<1.18$~T), the magnetic field and the plunger gate periods are, respectively, $\Delta B\approx17.0$~mT and $\Delta V_{PG}\approx18.0$~mV. The magnetic field period is close to that extracted at $\nu=1$ in the compressible regime, as expected for Aharonov-Bohm interference. The plunger gate period for the inner mode is $\Delta V_{PG}=18.0$~mV, approximately twice the period observed at $\nu=1$, consistent with the reduction in magnetic field needed to reach $\nu=2$~\cite{Jimmy19}. Using $\partial A / \partial V_{PG}$=0.198~$\mu$m$^2$V$^{-1}$ and the effective area $A=0.26$~$\mu$m$^2$ extracted at $\nu=1$, the expected period for pure Aharonov-Bohm interference at $\nu=2$ is calculated as $\Delta B=\phi_0/A=15.9$~mT and $\Delta V_{PG}=\phi_0/(B \partial A/ \partial V_{PG})=18.1$~mV, which align well with our experimentally observed periods. The observation of phase modulations in the low-field data suggests a compressible regime where the number of localized electrons in the interferometer varies with magnetic field and gate voltage. 

Near the center of the $\nu=2$ plateau (1.18~T$<B<1.28$~T), the periods change and a small number of discrete phase slips are observed, indicative of an incompressible regime in which disorder lowers the energy of a few discrete localized electronic states. The magnetic field period is larger in this regime, revealing the influence of bulk-edge coupling. In the incompressible regime the magnetic field period is measured to be $\Delta B\approx 32.9$~mT and the plunger gate period is $\Delta V_g \approx 15.0$~mV. Predominantly Aharonov-Bohm interference is maintained as the system transitions to the incompressible regime. At fields above $B=1.30$~T, the interference changes dramatically, suggesting a transition to a highly compressible regime and increased decoherence for the inner edge mode. This regime is not explored further in this study. 

Setting the QPCs to achieve total conductance $G=0.8e^2/h$ induces total reflection of the inner edge mode and partial transmission of the outer mode. As seen in Fig.~\ref{Fig:3}\textcolor{blue}{d}, at magnetic fields below $B=1.21$~T, the interference pattern exhibits lines of constant phase with negative slope with prominent modulations, reminiscent of the observations on the high magnetic field side of $\nu=1$. These additional modulations observed along lines of otherwise constant phase are consistent with periodic changes in the localized electron number in the interior of the interferometer. These phase modulations are only visible in the integer quantum Hall regime due to finite bulk-edge coupling; otherwise the phase change associated with the addition or removal of an electron is an unobservable factor of $2\pi$. 

For $B\leq1.21$~T, the interference periods are determined using a 2D Fast Fourier Transform (FFT) that yields $\Delta B\approx8.45$~mT and $\Delta V_{PG}\approx8.32$~mV as shown in Fig.~\ref{Fig:3}\textcolor{blue}{c}. Intriguingly, this magnetic field period is approximately half of the magnetic field period observed at $\nu=1$ in the compressible regime. This observation of $\phi_0/2$ periodicity is noteworthy as it is in conflict with naive expectations for interferometers in the Aharonov-Bohm limit~\cite{Halperin07,Halperin2011,Keyserlingk15,Frigeri2019,Rosenow2020,Feldman21}. The observation of $\phi_0/2$ periodicity when partially backscattering the outermost edge mode has previously been attributed to interference of charge $e^*=2$ quasiparticles in the integer quantum Hall regime in experiments using AlGaAs/GaAs interferometers~\cite{Choi15,sivan16,Sivan18}. Although the $\phi_0/2$ period observed in our device is consistent with observations in previous experimental works~\cite{Choi15,sivan16,Sivan18,yang2023,werkmeister2024}, it {\it does not} provide evidence for charge $e^*=2$ excitations. As we will demonstrate, our data are consistent with interference of charge $e^*=1$ electrons in the integer quantum Hall regime. 

The transition to an incompressible region near the center of the plateau at $B=1.25$~T is seen in the data of Fig.~\ref{Fig:3}\textcolor{blue}{d}. In this incompressible regime, $\Delta B\approx12.5$~mT and $\Delta V_{PG}\approx6.50$~mV. Interestingly, the magnetic field period is smaller than that observed at $\nu=1$ in the incompressible regime. In the interferometer studied here, the combined effects of bulk-edge coupling and edge-edge coupling associated with the fully reflected inner mode modify the periodicity in the incompressible regime at $\nu=2$. These coupling parameters can be determined {\it in operando} and constrain the possible value of the interfering charge. As seen in Fig.~\ref{Fig:3}\textcolor{blue}{d}, a few discrete phase slips are evident in the incompressible regime at $\nu=2$ for interference of the outer edge mode. The average magnitude of these phase slips is measured to be $\frac{\overline{\Delta \theta}}{2\pi}\approx0.38$, as shown in Fig.~\ref{Fig:3}\textcolor{blue}{e}, and is described in greater detail in the Supplementary Information. Referring to Eqn.~\ref{equation total phase variation for edge-edge coupling}, we may now consider the discontinuous change in interference phase associated with the removal or addition of a localized charged excitation while keeping all other parameters fixed such that $\delta (\bar A B)/\phi_0=0$, $\delta V_{PG}=0$, and $\delta N_2=0$. Under these circumstances, the expression for the change in phase $\Delta \theta$ associated with removal of a localized excitation in the interior of the interferometer reduces to:
\begin{eqnarray} \label{phase shift add or remove one charge}
\frac{\Delta\theta}{2\pi}=\frac{{e^*}^2 K_{1L}}{K_1}
\end{eqnarray}

The magnitude of the observed phase slips is determined by the effective charge and the coupling parameters $K_{1L}$ and $K_{1}$. To determine the effective interference charge experimentally, we must determine $K_{1L}$ and $K_1$. We utilize the model detailed in Refs.~\cite{Jimmy22,Jimmy2023} to determine $K_{1L}$, and $K_1$ from the measurements. It should be noted that the {\it sign} of the phase slips observed in our experiment is opposite to those observed in \cite{werkmeister2024}. The positive phase slips observed here suggest that the phenomenon is attributable to a change in the occupation of localized excitations in the interior of the interferometer.

In order to relate $K_1$ and $K_{1L}$ to experimentally measurable quantities, we note that the total energy of the interferometer, $E_{tot}$, may be written as a combination of electron-electron interaction energy, $E_{int}$, and the single-particle energy, $E_{sp}$, due to electrostatic confinement in the interferometer: $E_{tot}=E_{int}+E_{sp}$. To determine $E_{int}$, we model the interferometer as a quantum dot. The energy of the electron-electron interaction is then given by $E_{int}=(e\delta q_{tot})^2/2C_0$, where $e\delta q_{tot}=e(\delta q_1+\delta q_2+\delta q_b)$ is the total charge within the interferometer, and $C_0$ is the self-capacitance of the interferometer. The interaction energy can be written as: 
\begin{eqnarray}
E_{int}=&&(e^2/{C_0})(\delta q_1^2/{2}+\delta q_2^2/{2}+\delta q_b^2/{2}+\nonumber\\
&&\delta q_1 \delta q_2+\delta q_1 \delta q_b+\delta q_2 \delta q_b)   
\end{eqnarray}
As illustrated in Fig.~\ref{Fig:2}\textcolor{blue}{a}, the measurement of the differential conductance in the Coulomb blockade regime at $B=0$~T produces a charging energy of $e^2/C\approx95$~$\mu$eV. We can refine our determination of $C_0$ by subtracting the contribution from the single particle level spacing at $B=0$~T determined by the finite density of states in a 2D system~\cite{Jimmy22}. For a device with an area $A\approx0.26$~$\mu$m$^2$, the contribution of the quantum capacitance is $e^2/C_{quantum}=\pi \hbar^2/(m^*A)\approx11$~$\mu$eV. This yields $e^2/C_0=e^2/C-e^2/C_{quantum}=95$~$\mu$eV-11~$\mu$eV$=84$~$\mu$eV. 

The single-particle energy, $E_{sp}$, is the energy cost to add charge to each edge mode in the presence of the external electrostatic confining potential when the area of the interferometer changes. This energy is given by $E_{sp}=e\delta q^2_i \phi_0 v_i/{2L}$, where $L$ is the perimeter of the interference path calculated from the effective area $A$ as $L\approx4\sqrt{A}\approx2.04$~$\mu$m, and $v_i$ is the velocity of the $i^{th}$ edge state~\cite{Jimmy22}. The edge velocity is experimentally determined by the node spacing, $\Delta V_{SD}$, observed in finite DC source-drain bias measurements of the interference signal, expressed as $\Delta V_{SD}=hv_i/{(ee^*L)}$~\cite{chamon97,Sahasrabudhe2018,Jimmy22,Jimmy2023,Feldman24}. At $\nu=2$, two edge modes contribute to the single-particle energy such that $E_{sp}=\frac{e^*\delta q_1^2}{2}e\Delta V_{SD1}+\frac{e^*\delta q_2^2}{2}e\Delta V_{SD2}$. $\Delta V_{SD1}$ and $\Delta V_{SD2}$ are the node spacings determined by finite bias measurements for the outer mode and inner mode, respectively. 

Combining Eqn.~\ref{total energy} with the expressions for the electron-electron interaction energy and the single particle energy allows us to write expressions for the bulk-edge coupling, $K_{1L}$, and the stiffness of the edge mode, $K_1$, in terms of experimentally measurable quantities~\cite{Jimmy22, Jimmy2023}:

\begin{equation}
K_{1L}=\frac{e^2}{C_0}\label{equation bulk-edge coupling}
\end{equation}
\begin{equation}
K_{1}=\frac{e^2}{C_0}+{e^*}e\Delta V_{SD1}\label{equation edge stiffness}
\end{equation}
Eqn.~\ref{equation bulk-edge coupling} yields $K_{1L}=e^2/C_0=84$~$\mu$eV. In the integer quantum Hall regime, the edge stiffness $K_1$ can be determined from the measurement of the charging energy at $B=0$~T and the measurement of the spacing of the nodes with finite source-drain bias as shown in Fig.~\ref{Fig:3}\textcolor{blue}{f} for the outer mode at $\nu=2$. The FFT amplitude versus $V_{SD}$ is shown in Fig.~\ref{Fig:3}\textcolor{blue}{g} which yields $\Delta V_{SD1}=132$~$\mu$V. For our analysis of $\phi_0/2$-periodic interference for the outer mode at $\nu=2$, we have retained the magnitude of interfering charge as a variable to be determined in experiment. In this case, the identification of an incompressible regime punctuated by few discrete phase slips for interference of the outer mode at $\nu=2$ provides crucial information. We may combine Eqn.~\ref{phase shift add or remove one charge} with our expressions for $K_{1L}$ and $K_1$ (Eqn.~\ref{equation bulk-edge coupling} and Eqn.~\ref{equation edge stiffness}) into a simple quadratic equation for the interfering charge in terms of the measured average phase slip $\overline{\Delta \theta}$ and the experimentally determined values for $K_1$ and $K_{1L}$:

\begin{eqnarray} \label{Equation to determine charge at v=2 part 1}
\frac{\overline{\Delta \theta}}{2\pi}=\frac{{e^*}^2 K_{1L}}{K_1}= \frac{{e^*}^2\frac{e^2}{C_0}}{e^2/C_0+ee^*\Delta V_{SD1}}
\end{eqnarray}
which may be combined with the measured values $e\Delta V_{SD1}=132\mu$eV and $e^2/C_0=84\mu$eV to yield,
\begin{eqnarray} \label{Equation to determine charge at v=2}
\frac{\overline{\Delta \theta}}{2\pi}=0.38=\frac{{e^*}^2 84 \mu eV}{84 \mu e V+e^*132 \mu e V}
\end{eqnarray}
Solving Eqn.~\ref{Equation to determine charge at v=2} yields an interfering charge $e^*=0.98$ and therefore we conclude $e^*=1$ for the outer edge mode at $\nu=2$. Hence, our {\it in operando} determination of all relevant coupling parameters directly indicates that the interfering charge is $e^*=1$ for the outer mode at $\nu=2$.

To complete our characterization of all coupling constants, we need to determine $K_{12}$, the edge-edge coupling strength at $\nu=2$. This may be achieved through examination of Eqn.~\textcolor{blue}{10} and measurement of the magnetic field period $\Delta B\approx12.5$~mT in the incompressible regime. The interferometer area $A=0.26\mu$m$^2$ and the interference charge $e^*=1$. The edge stiffness is $K_1=216\mu$eV, while $K_{1L}$ was determined to be $84\mu$eV. Therefore, $K_{1L}/K_1\approx0.39$. These measured parameters may be used with Eqn.~\textcolor{blue}{10} to calculate $K_{12}/K_1$. 
\begin{equation} \label{Bperiod in the incompressible region for edge-edge coupling inverted}
\frac{K_{12}}{K_1}=\frac{\phi_0}{e^*A\Delta B} + 2\frac{K_{1L}}{K_1}-1
\end{equation}
Equation 18 yields $K_{12}/K_1\approx1$. The strong edge-edge coupling derived here is consistent with the observation of periodicity $\phi_0/2$ in the compressible regime at $\nu=2$. Examination of Eqn.~\textcolor{blue}{12} indicates that the magnetic field period becomes $\Delta B=\frac{\phi_0}{2\bar A}$ when $K_{12}/K_1=1$. Our analysis at $\nu=2$ demonstrates that the interfering charge is $e^*=1$ and strong capacitive edge-edge coupling generates periodicity $\phi_0/2$ in the absence of an exotic electron pairing mechanism. The tunable bulk-edge and edge-edge coupling strengths afforded by the screening well heterostructure design facilitate quantitative analysis.

\subsection{Interference at \texorpdfstring{$\nu=3$}{Lg}}\label{subsec:level4}
\begin{figure*}
\centering
\includegraphics[width=1\textwidth]{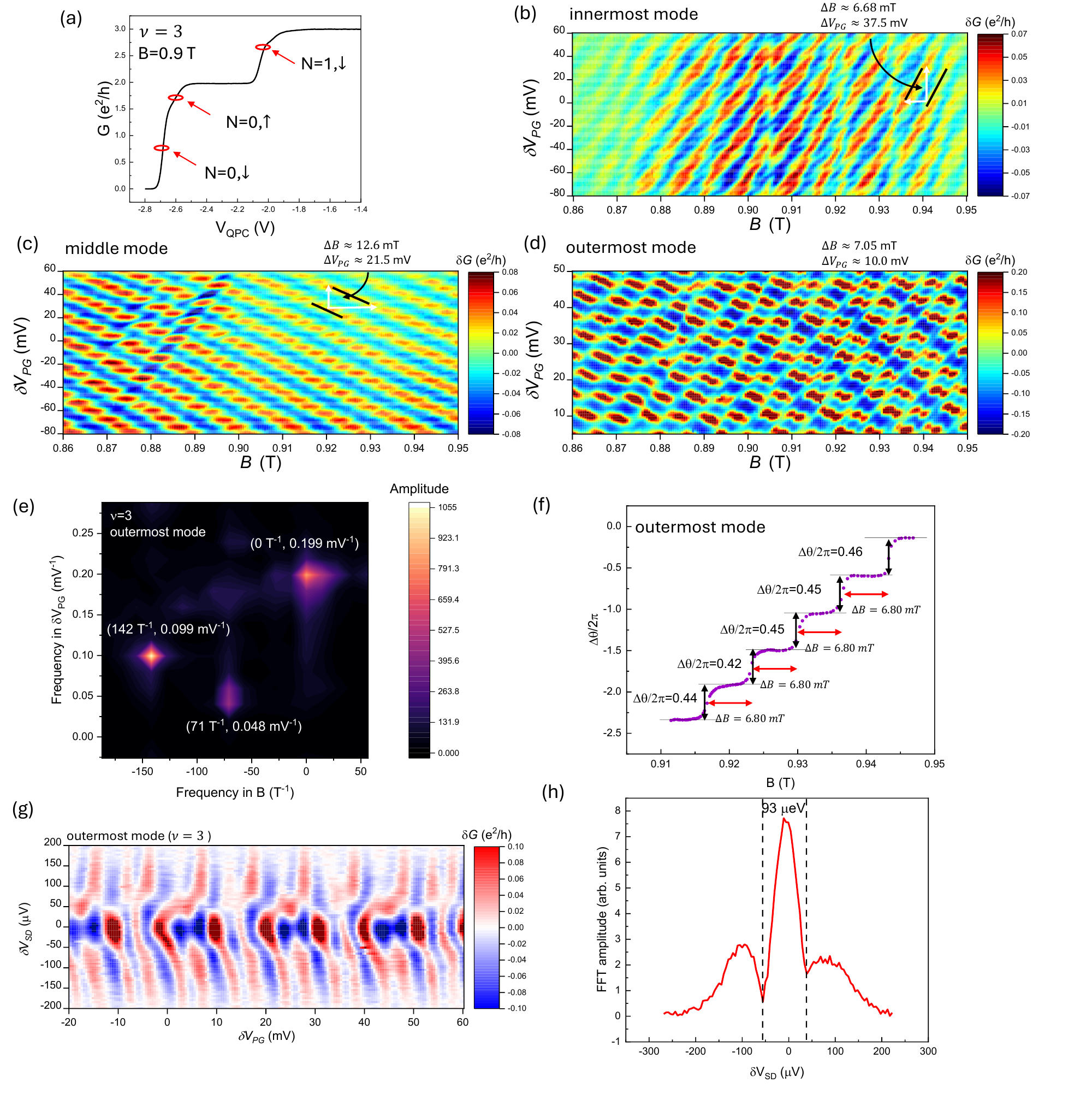}
\caption{\label{Fig:4} \textbf{Interference at $\nu=3$} (a) QPC conductance versus gate voltage at $\nu=3$. $G=3\frac{e^2}{h}$ and $G=2\frac{e^2}{h}$ plateaus are observed. The $G=\frac{e^2}{h}$ plateau is not well-formed, likely due to the small N=0 Landau level spin gap at $B=0.9$~T, which prevents formation of a wide incompressible region between the spin-down and spin-up N=0 Landau level edge modes. However, an inflection near $G=\frac{e^2}{h}$ is observed, indicating a transition from backscattering the spin-down N=0 edge mode to backscattering the spin-up N=0 edge mode. Red circles indicate the transmission used to interfere each mode at $\nu=3$. (b) Interference generated by partially backscattering the innermost edge mode (spin-down N=1 Landau level). (c) Interference produced when backscattering the middle edge mode (spin-up N=0 Landau level). Black solid lines are drawn along lines of nearly constant phase in panels (b), and (c). (d) Interference when partially backscattering the outermost edge mode (spin-down N=0 Landau level). (e) 2D FFT for interference of the outermost edge mode at $\nu=3$ identifying the most significant frequencies in the interference. (f) Plot of the magnitude of discontinuous phase slips observed in Fig.~\ref{Fig:4}d with $\overline{\Delta\theta}/2\pi\approx0.44$. (g) Differential conductance measurement for the outermost mode at $\nu=3$. (h) Plot of FFT amplitude vs. $\delta V_{SD1}$ used to determine the node spacing for the outermost mode.
}
\end{figure*}

At $\nu=3$, we sequentially set the transmission of the QPCs to $G=2.8e^2/h$, $1.8e^2/h$, and $0.6e^2/h$ to selectively interfere the innermost, the middle, and the outermost edge modes as illustrated in Fig.~\ref{Fig:4}\textcolor{blue}{a}. As shown in Fig.~\ref{Fig:4}\textcolor{blue}{b}, for the innermost edge mode, constant phase lines within the $B-\delta V_{PG}$ plane exhibit a positive slope, a characteristic of interference in the Coulomb-dominated regime~\cite{Halperin07,Halperin2011,Keyserlingk15,Frigeri2019,Feldman21}. For the innermost mode, the magnetic field period is 6.68~mT and the plunger gate voltage period is 37.5~mV. The strength of the bulk-edge coupling $K_{3L}$ and the edge stiffness $K_3$ are estimated from the $B=0$~T Coulomb blockade data and finite DC source-drain bias measurements at $\nu=3$ using the model developed in~\cite{Jimmy22,Jimmy2023}. We extract the ratio $K_{3L}/K_3\approx0.58$ for this edge mode. Given that $K_{3L}/K_3>0.5$, the interference for the innermost mode occurs in the Coulomb-dominated regime as theoretically described in~\cite{Halperin2011,Keyserlingk15,Frigeri2019,Feldman21}. This data provides an example of how an interferometer can switch the Aharonov-Bohm regime to the Coulomb-dominated regime depending on device tuning.

When the innermost edge is fully reflected and the middle edge mode is weakly backscattered, Aharonov-Bohm-like interference with lines of nearly constant phase displaying negative slope is observed, as shown in Fig.~\ref{Fig:4}\textcolor{blue}{c}. The periods of this interference are extracted using a 2D FFT. The measured magnetic field period is $\Delta B\approx12.6$~mT, and the plunger gate voltage period is $\Delta V_{PG}\approx21.5$~mV. Note that the magnetic field period is approximately twice that measured for the innermost mode, consistent with theoretical expectations as the number of fully transmitted modes has now decreased to one~\cite{Halperin2011,Keyserlingk15,Frigeri2019,Feldman21} while maintaining fixed filling factor in the interior of the interferometer. For the middle mode, $K_{2L}/K_2\approx0.37$ is determined by analyzing finite DC bias interference and Coulomb blockade spectroscopy at $B=0$~T, indicating a reduction in the ratio of bulk-edge coupling to edge stiffness compared to the innermost mode. This reduction is consistent with the transition from Coulomb-dominated interference for the innermost edge mode to a predominantly Aharonov-Bohm-like pattern for the middle edge mode. Additional amplitude modulations along the negatively sloped lines of constant phase are also clearly observed. The occurrence of these modulations for Aharonov-Bohm-like interference is associated with moderate bulk-edge coupling in a compressible regime in an integer quantum Hall state~\cite{Halperin2011,Keyserlingk15,Frigeri2019,Feldman21}. 

It is interesting to note that, unlike $\nu=1$ and $\nu=2$, the interference pattern throughout the $\nu=3$ plateau appears to reflect only a compressible regime of operation. There are no visible changes in periodicity that are normally associated with the transition from a compressible state on the flanks of the quantum Hall plateau to an incompressible state near the center of the quantum Hall plateau. The moderately strong Coulomb coupling and the smaller bulk excitation gap at $\nu=3$ at $B=0.9$~T in this instance make it energetically favorable to keep the total charge in the interior of the interferometer fixed rather than to keep the filling factor fixed~\cite{Rosenow2020}. The magnetic field range of an incompressible region is expected to be proportional to the ratio of the quantum Hall state energy gap to the charging energy required to add one electron to the interferometer, as described in~\cite{Rosenow2020} and confirmed by previous experimental observations~\cite{Jimmy20,Jimmy22,Jimmy2023}. One of our previous experiments reported an incompressible regime when partially backscattering the outermost edge mode at $\nu=3$~\cite{Jimmy19}. In that interferometer, the charging energy was measured to be $e^2/C\approx17$~$\mu$eV~\cite{Jimmy19}. In our current experiment, the interferometer has a significantly higher charging energy of $e^2/C\approx95$~$\mu$eV. Furthermore, the effective energy gap at $\nu=3$ is reduced because of the significantly lower electron density used here compared to the value used in~\cite{Jimmy19}. The reduced excitation gap-to-charging-energy ratio likely results in the absence of a detectable incompressible region at $\nu=3$ in the current experiment.

When interfering the outermost edge mode and isolating the two inner modes from the ohmic leads, a complex lattice-like pattern of interference develops, as seen in Fig.~\ref{Fig:4}\textcolor{blue}{d}. Dramatic phase slips, interspersed within short stretches of continuous phase evolution, are evident. The periodicities for this configuration are determined via 2D FFT, as shown in Fig.~\ref{Fig:4}\textcolor{blue}{e}. The strongest component in the 2D FFT decomposition corresponds to a magnetic field period of $\Delta B\approx7.05$~mT and a gate voltage period of $\Delta V_{PG}\approx10.0$~mV. Similar to the outer mode at $\nu=2$ in the compressible regime, this magnetic field period is consistent with an approximate flux periodicity $\phi_0/2$ for the outermost edge mode at $\nu=3$. The lattice-like interference pattern represents an admixture of Aharonov-Bohm and Coulomb-dominated oscillations, consistent with the theoretical predictions for moderate bulk-edge coupling and strong edge-edge coupling~\cite{Halperin2011,Keyserlingk15,Frigeri2019,Feldman21}. The observation of short stretches of constant phase evolution interrupted by periodic phase slips indicates that the interference occurs in a compressible regime. As was done at $\nu=2$, we can quantify the magnitude of the discrete phase slips as shown in Fig.~\ref{Fig:4}\textcolor{blue}{f} with $\overline{\Delta \theta}/{2\pi}=0.44$.

As was done at $\nu=2$, we must determine the interfering charge for the outermost edge at $\nu=3$ that gives rise to the approximate periodicity $\phi_0/2$. $K_{1L}=84~\mu$eV has already been determined. The differential conductance measurements of interference for the outermost mode at $\nu=3$ are shown in Fig.~\ref{Fig:4}\textcolor{blue}{g}, resulting in the node spacing $\Delta V_{SD1}=93~\mu$eV shown in Fig.~\ref{Fig:4}\textcolor{blue}{h}. The quadratic equation relating the average value of the measured phase slips with the effective charge now reads: 
\begin{equation}
\overline{\Delta \theta}/2\pi =0.44=\frac{{e^*}^2 84\mu eV}{84\mu eV + e^* 93\mu eV}    
\end{equation}
The interfering charge for the outermost mode at $\nu=3$ is also $e^*=1$. Once the effective charge is specified, the ratio $K_{1L}/K_1\approx0.47$ is determined.

Unlike $\nu=2$, where we observed interference in both the incompressible and compressible regimes, interference occurs only in the compressible regime at $\nu=3$. As discussed in the Introduction, we can evaluate $K_{21}/K_1$ using Eqn.~\textcolor{blue}{12}. 
Using $\bar{A}=0.26~\mu$m$^2$ and the magnetic field period extracted from the FFT of the data in Fig.~\ref{Fig:4}\textcolor{blue}{d}, $\Delta B=7.05$~mT, yields $K_{12}/K_1\approx1$. The extracted edge-edge coupling ratio of $K_{12}/K_1\approx1$ indicates a strong interaction between the outermost interfering edge and the isolated inner edges. This strong capacitive edge-edge coupling is consistent with the observed approximately $\phi_0/2$ flux periodicity. The specification of the coupling parameters for interference at $\nu=3$ is complete. Again, the analysis indicates that an interference charge of $e^*=1$ and strong edge-edge coupling are sufficient to generate periodicity $\phi_0/2$.

\section{Discussion}\label{sec:level3}
Recent developments in Fabry-P{\'e}rot interferometry experiments~\cite{Choi15,sivan16,Sivan18,yang2023,werkmeister2024} and theory~\cite{Frigeri2019,Frigeri2020} have drawn attention to the origin of the flux period $\phi_0/2$ in the integer quantum Hall regime when interfering the outermost edge mode. In previous experiments using GaAs/AlGaAs 2DEGs, the observed $\phi_0/2$ flux period in the integer quantum Hall regime was attributed to electron pairing \cite{Choi15,sivan16,Sivan18}. A recent experiment in a graphene interferometer~\cite{werkmeister2024} explains the transition in periodicity from $\phi_0$ to $\phi_0/2$ in the integer quantum Hall regime as a result of edge-edge coupling and single-electron interference without invoking an electron pairing mechanism. In this graphene experiment, the importance of edge-edge capacitive coupling is highlighted; our results in an AlGaAs/GaAs interferometer substantiate the conclusions of \cite{werkmeister2024} with the quantification of the edge-edge coupling parameter in the regime in which $\phi_0/2$ periodicity is observed. In this regard, our data and analysis support the conclusions of Ref.~\cite{werkmeister2024}. However, in the graphene interferometer, the magnitude of the interfering charge could not be specified in the strong edge-edge coupling limit because discrete phase slips were not observed. The ability to determine the interfering charge through {\it in operando} extraction of all the necessary coupling parameters in our experiment removes ambiguity. To amplify this point, another recent experiment in graphene by a different group has claimed evidence for even more exotic pairing behavior involving electronic triplets~\cite{yang2023} in the integer quantum Hall regime, adding a new and confounding twist to this evolving story of electron correlations in interferometers operated in the integer quantum Hall regime. In a device with minimal bulk-edge coupling and strong edge-edge coupling, it is challenging to discriminate between the effects of electron pairing, however improbable, and edge-edge capacitive coupling. The introduction of moderate bulk-edge coupling in our AlGaAs/GaAs interferometer allows the disambiguation of these two phenomena by examining the magnitude of phase slips at $\nu=2$ and $\nu=3$ in the strong edge-edge coupling limit.

It is important to note that an assumption of an interfering charge $e^*=2$ on the outer edge at $\nu=2$ and $\nu=3$ is not consistent with our experimental results. We have intentionally studied a structure that simultaneously possesses finite bulk-edge coupling and edge-edge coupling to explore the impact of each mechanism and explore how control of these couplings may be used to determine the charge of the interfering particles. The observation of phase slips with magnitude that is a precise, but non-integer, multiple of $2\pi$, at $\nu=2$ and $\nu=3$ is a key feature of our experiments that allows us to determine the charge of the interfering particles in the integer quantum Hall states. Our observations are clearly in variance with a model of interference with the charge $e^*=2$, and strongly indicate that the interfering particles carry the charge of a single electron. The internal consistency of multiple independent measurements and the minimal capacitive model of interferometer energetics that can explain multiple experiments conducted with several interferometers \cite{Jimmy20,Jimmy22,Jimmy19} support a picture in which $\phi_0/2$ flux periodicity in the integer quantum Hall regime is generated by edge-edge coupling and the interfering charge is a singly charged electron. 

To conclude, we studied a device characterized by moderate bulk-edge coupling and strong edge-edge coupling, enabling a quantitative assessment of the impact of these coupling mechanisms on the flux periodicity of interference. We determine the charge of interfering particles in the integer quantum Hall regime using the magnitude of discrete phase slips associated with the introduction of localized charged excitations in the interior of the interferometer. Using a simple capacitive model, we analyze the interference phenomena for $1\leq\nu \leq 3$, providing information on the underlying source of the different frequencies observed in Fabry-P{\'e}rot interferometers with multiple edge modes. The interfering charge is determined to be $e^*=1$.

\section{Methods}\label{sec:level5}
Mesas were defined by optical lithography and 50:5:1 H$_2$O:H$_3$PO$_4$:H$_2$O$_2$ wet etching. 8~nm~Ni/~80~nm~Ge/~160~nm~Au/~36~nm~Ni ohmic contacts were deposited and annealed to make electrical contact with the 2DEG. Electron beam lithography and electron beam evaporation of 5~nm~Ti/10~nm~Au were used to define the interferometer gates. Optical lithography and electron beam evaporation of 20~nm~Ti/150~nm~Au were used to define bond pads and surface gates in the vicinity of ohmic contacts. The substrate was thinned to approximately 70~$\mu$m, and metallic back gates were defined to deplete the 2DEG in the bottom screening well in the vicinity of the ohmic contacts. The back gates were patterned by optical lithography and 100~nm~Ti/~150~nm~Au was deposited by electron beam evaporation.

Conductance measurements were made using standard lock-in amplifier techniques, employing an excitation voltage V$_{ex}$$\leq$20~$\mu$V at a frequency of 43~Hz. These measurements were carried out in a dilution refrigerator at a mixing chamber temperature of $T=10$~mK, unless otherwise specified. All data were obtained with the plunger gate biased at -0.7~V. For finite DC bias measurements, a 5~$\mu$V AC voltage was typically used, and the DC source-drain voltage was swept using a programmable digital-to-analog converter.

\section{Data availability}\label{sec:level6}
The data supporting the figures in this manuscript are available at https://doi.org/10.5281/zenodo.15252948. All other data that support the findings of this study are available from the corresponding authors upon request.

\section{Code availability}\label{sec:level7}
The code used to analyze the data presented in this manuscript is available at https://doi.org/10.5281/zenodo.15252948

\providecommand{\noopsort}[1]{}\providecommand{\singleletter}[1]{#1}%

\section{Acknowledgements}\label{sec:level8}
This research is sponsored by the U.S. Department of Energy, Office of Science, Office of Basic Energy Sciences, under award number DE-SC0020138. The content of the information presented here does not necessarily reflect the position or the policy of the US government, and no official endorsement should be inferred.
M. J. Manfra acknowledges D. E. Feldman and B. Rosenow for insightful comments on a preliminary version of this manuscript. M. J. Manfra thanks Bertrand Halperin for several suggestions which significantly improved the logic and rigor of our manuscript.  

\section{Author contributions}\label{sec:level9}
S.L., and M.J.M. wrote the manuscript. J.N. and M.J.M. designed the heterostructures. S.L. and G.G. conducted molecular beam epitaxy growth. J.N. fabricated the devices. S.L. performed the measurements and analyzed the data with input from M.J.M.

\section{Competing interests}\label{sec:level10}
The authors declare no competing interests.

\clearpage

\end{document}